\newcommand{\Planck}{\textit{Planck }}
\newcommand{\Herschel}{\textit{Herschel }}
\newcommand{\blastpol}{BLASTPol}

\newcommand{\um}{$\mu$m}                            
\newcommand{\av}{$A_{\mathrm{V}}$}

\def\pratone{$p_{250}/p_{350}$}
\def\prattwo{$p_{500}/p_{350}$}
\def\pratthree{$p_{850}/p_{350}$}

\def\conslinearb{$\phantom{-}1.3\pm2.7$}
\def\conslineara{$1.03\pm0.06$}
\def\conspowerlawb{$\phantom{-}0.06\pm0.13$}
\def\conspowerlawa{$1.03\pm0.06$}
\def\conspolyb{$0.7\pm1.4$}
\def\conspolyc{$-1.8\pm6.3$}
\def\conspolya{$1.04\pm0.04$}

\def\agglinearb{$\phantom{-}0.2\pm3.1$}
\def\agglineara{$0.99\pm0.03$}
\def\aggpowerlawb{$-0.01\pm0.14$}
\def\aggpowerlawa{$0.98\pm0.02$}
\def\aggpolyb{$1.7\pm1.3$}
\def\aggpolyc{$-6.6\pm5.2$}
\def\aggpolya{$1.01\pm0.03$}

\def\intlinearb{$\phantom{-}0.9\pm2.8$}
\def\intlineara{$1.01\pm0.04$}
\def\intpowerlawb{$\phantom{-}0.03\pm0.13$}
\def\intpowerlawa{$1.01\pm0.04$}
\def\intpolyb{$1.1\pm1.1$}
\def\intpolyc{$-3.7\pm5.0$}
\def\intpolya{$1.03\pm0.03$}

\def\cloudlinearb{$-2.9\pm3.5$}
\def\cloudpowerlawb{$-0.21\pm0.13$}
\def\cloudpolyb{$2.3\pm0.6$}
\def\cloudpolyc{$-8.5\pm2.5$}

\def\inlauralinearb{$-1.0\pm2.9\phantom{0}$}
\def\inlaurapowerlawb{$-0.09\pm0.12\phantom{0}\phantom{0}$}
\def\inlaurapolyb{$2.3\pm1.3$}
\def\inlaurapolyc{$-12\pm6.7\phantom{0}$}

\def\outlauralinearb{$1.0\pm2.8$}
\def\outlaurapowerlawb{$0.037\pm0.13\phantom{0}$}
\def\outlaurapolyb{$1.1\pm1.1$}
\def\outlaurapolyc{$-3.6\pm4.9\phantom{0}$}

\def\deg{\ifmmode^\circ\else$^\circ$\fi}
\def\arcm{\ifmmode {^{\scriptstyle\prime}}
          \else $^{\scriptstyle\prime}$\fi}

\newcommand{\hnatalie}[1]{}
\documentclass[iop,numberedappendix,twocolappendix]{emulateapj}
\usepackage{float}
\usepackage[morefloats=7]{morefloats}
\usepackage[breaklinks,colorlinks,citecolor=blue]{hyperref} 
\usepackage[all]{hypcap}
\usepackage{booktabs}

\slugcomment{In preparation for submission to ApJ}

\providecommand{\sorthelp}[1]{}

\shorttitle{Vela C Polarization Spectrum}
\shortauthors{Gandilo et al.}

\begin{document}

\title{Submillimeter Polarization Spectrum in the Vela C Molecular Cloud}
\author{Natalie N. Gandilo\altaffilmark{1,2},
Peter A. R. Ade\altaffilmark{3},
Francesco E. Angil\`e\altaffilmark{4},
Peter Ashton\altaffilmark{5},
Steven J. Benton\altaffilmark{6,7},
Mark J. Devlin\altaffilmark{4},
Bradley Dober\altaffilmark{4},
Laura M. Fissel\altaffilmark{5},
Yasuo Fukui\altaffilmark{8},
Nicholas Galitzki\altaffilmark{4},
Jeffrey Klein\altaffilmark{4},
Andrei L. Korotkov\altaffilmark{9},
Zhi-Yun Li\altaffilmark{10},
Peter G. Martin\altaffilmark{11},
Tristan G. Matthews\altaffilmark{5},
Lorenzo Moncelsi\altaffilmark{12},
Fumitaka Nakamura\altaffilmark{13},
Calvin B. Netterfield\altaffilmark{2,7},
Giles Novak\altaffilmark{5},
Enzo Pascale\altaffilmark{3},
Fr{\'e}d{\'e}rick Poidevin\altaffilmark{14,15},
Fabio P. Santos\altaffilmark{5},
Giorgio Savini\altaffilmark{16},
Douglas Scott\altaffilmark{17},
Jamil A. Shariff\altaffilmark{2,18},
Juan Diego Soler\altaffilmark{19},
Nicholas E. Thomas\altaffilmark{20},
Carole E. Tucker\altaffilmark{3},
Gregory S. Tucker\altaffilmark{10},
Derek Ward-Thompson\altaffilmark{21}}

\altaffiltext{1}{Department of Astronomy \& Astrophysics, University of Toronto, 50 St. George Street Toronto, ON M5S 3H4, Canada}
\altaffiltext{2}{Department of Physics and Astronomy, Johns Hopkins University, 3701 San Martin Drive, Baltimore, Maryland, USA}
\altaffiltext{3}{Cardiff University, School of Physics \& Astronomy, Queens Buildings, The Parade, Cardiff, CF24 3AA, UK}
\altaffiltext{4}{Department of Physics \& Astronomy, University of Pennsylvania, 209 South 33rd Street, Philadelphia, PA, 19104, US.A.}
\altaffiltext{5}{Center for Interdisciplinary Exploration and Research in Astrophysics (CIERA) and Department\ of Physics \& Astronomy, Northwestern University, 2145 Sheridan Road, Evanston, IL 60208, USA}
\altaffiltext{6}{Department of Physics, Princeton University, Jadwin Hall, Princeton, NJ 08544, USA}
\altaffiltext{7}{Department of Physics, University of Toronto, 60 St. George Street Toronto, ON M5S 1A7, Canada}
\altaffiltext{8}{Department of Physics and Astrophysics, Nagoya University, Nagoya 464-8602, Japan}
\altaffiltext{9}{Department of Physics, Brown University, 182 Hope Street, Providence, RI, 02912, USA}
\altaffiltext{10}{Department of Astronomy, University of Virginia, 530 McCormick Rd, Charlottesville, VA 22904, USA}
\altaffiltext{11}{CITA, University of Toronto, 60 St. George Street, Toronto, ON M5S 3H8, Canada}
\altaffiltext{12}{California Institute of Technology, 1200 E. California Boulevard, Pasadena, CA, 91125, USA}
\altaffiltext{13}{National Astronomical Observatory, Mitaka, Tokyo 181-8588, Japan}
\altaffiltext{14}{Instituto de Astrofísica de Canarias, E-38200 La Laguna, Tenerife, Spain}
\altaffiltext{15}{Universidad de La Laguna, Dept. Astrof\'{i}sica, E-38206 La Laguna, Tenerife, Spain}
\altaffiltext{16}{Department of Physics \& Astronomy, University College London, Gower Street, London, WC1E 6BT, UK}
\altaffiltext{17}{Department of Physics \& Astronomy, University of British Columbia, 6224 Agricultural Road, Vancouver, BC V6T 1Z1, Canada}
\altaffiltext{18}{Department of Physics, Case Western Reserve University, 2076 Adelbert Road, Cleveland, OH, 44106, U.S.A}
\altaffiltext{19}{Institute d'Astrophysique Spatiale, CNRS (UMR8617) Universit\'{e} Paris-Sud 11, B\^{a}timent 121, Orsay, France}
\altaffiltext{20}{NASA/Goddard Space Flight Center, Greenbelt , MD 20771, USA}
\altaffiltext{21}{Jeremiah Horrocks Institute, University of Central Lancashire, PR1 2HE, UK}

\begin{abstract}
Polarization maps of the Vela C molecular cloud were obtained at 250, 350, and 500\,\um~during the 2012 flight of the balloon-borne telescope \blastpol. These measurements are used in conjunction with 850\,\um~data from \Planck to study the submillimeter spectrum of the polarization fraction for this cloud. The spectrum is relatively flat and does not exhibit a pronounced minimum at $\lambda \sim$~350\,\um~as suggested by previous measurements of other molecular clouds. The shape of the spectrum does not depend strongly on the radiative environment of the dust, as quantified by the column density or the dust temperature obtained from \Herschel data. The polarization ratios observed in Vela C are consistent with a model of a porous clumpy molecular cloud being uniformly heated by the interstellar radiation field.  \end{abstract}

\maketitle
\section{Introduction}\label{sec:Intro}

The role that magnetic fields play in the process of star formation is not well understood. This question can be addressed by observing the strength and morphology of magnetic fields in the dense molecular clouds where stars form. One important method of observing magnetic fields in star-forming regions is through submillimeter polarimetry. Because dust grains tend to align with their long axes perpendicular to the direction of the local magnetic field, the linearly polarized thermal emission from the dust grains can be used to trace the magnetic field direction in the plane of the sky (see \citet{Lazreview} for a review).

The use of polarimetry to probe magnetic fields requires a good understanding of the mechanism by which dust grains align with magnetic fields and how the alignment depends on the local environment.
Multiwavelength observations of polarized dust emission can probe the conditions under which dust grains are aligned and test theories of grain alignment mechanisms. The theory of radiative torques \citep[RATs;][]{Dolginov1976,Lazarian2007} has become the most favored model for how dust grains align with magnetic fields. This model proposes that an irregularly shaped dust grain with a different cross-section for right- and left-handed photons will get spun up by unpolarized light in the presence of an anisotropic radiation field, ultimately causing the particle to rotate with its long axis perpendicular to magnetic field lines (see \citet{Andersson2015} for a review).

At visible and near infrared wavelengths, much has been inferred about the physical properties of dust grains by studying starlight polarization originating from dichroic extinction \citep{Whittet2001,Whittet2008}. In particular, large grains (radii $\geq0.1$\,\um) are much more efficiently aligned than small grains \citep{Kim1995}; amorphous silicate grains are better aligned than carbonaceous grains (including polycyclic aromatic hydrocarbons; \citealt{Smith2000,Chiar2006}; see \citealt{Draine2003}). The simplest models for molecular cloud dust \citep{Hildebrand1999,Bethell2007} predict submillimeter polarization spectra that are flat to 10-20\%. However, the observations to date \citep{Hildebrand1999,Vaillancourt2008,Vaillancourt2012,Zeng2013} show a polarization fraction with more variation, rising away from a minimum at 350\,\um~(see Figure \ref{allspec} below).

To explain the rise in the submillimeter spectrum a model in which the colder grains are better aligned than the warmer grains is needed. \citet{Bethell2007} have shown that this can be achieved by applying the RAT model to starless clouds. They model a clumpy molecular cloud structure in which external photons can penetrate deep into the cloud. These photons heat all grains, but the larger grains tend to be cooler, because they are more efficient emitters. At the same time, the mechanism for aligning larger grains is more efficient \citep{Cho2005}. Therefore, their model predicts that the cooler grains are better aligned and that the polarization spectrum rises with wavelength. However, the predicted submillimeter rise is much shallower than is seen in the published observations. The model also predicts a polarization spectrum rising with wavelength in the far-IR, and so it cannot explain the falling far-IR spectrum that has been observed.

The Balloon-borne Large Aperture Submillimeter Telescope for Polarimetry (\blastpol) observed the Vela C molecular cloud at 250, 350, and 500\,\um. Multiwavelength observations bracketing the 350\,\um~band provide a new opportunity to study the shape of the polarization spectrum. Vela C is a relatively nearby molecular cloud, spanning 35\,pc at a distance of $d\sim700$\,pc \citep{Liseau1992} and contains $5\,\times\,10^4\,$M$_{\sun}$ of dense gas \citep{Yamaguchi1999}. It includes a large number of objects in the early stages of star formation \citep{Netterfield2009} as well as a bright compact \ion{H}{2} region, RCW 36, thus providing an opportunity to study grain alignment in varying radiative conditions. \citet{Hill2011} analyzed Vela C using data from the \Herschel HOBYS program, and identified five sub-regions in the cloud, as defined at an \av\,$>$\,7\,mag threshold.

This paper presents polarization data in Vela C from \blastpol~at 250, 350, and 500\,\um, and from \Planck at 850\,\um. Section~\ref{sec:obs} describes the observations of submillimeter polarization in Vela C. Section~\ref{sec:results} presents the polarization spectrum of Vela C and investigates the effect of the radiative environment. Section~\ref{sec:discuss} discusses the implications of this work for theories of dust grain alignment and Section~\ref{sec:summ} summarizes the results.

\section{Observations and Data Analysis}
\label{sec:obs}
\begin{figure}
\epsscale{1.22}
\plotone{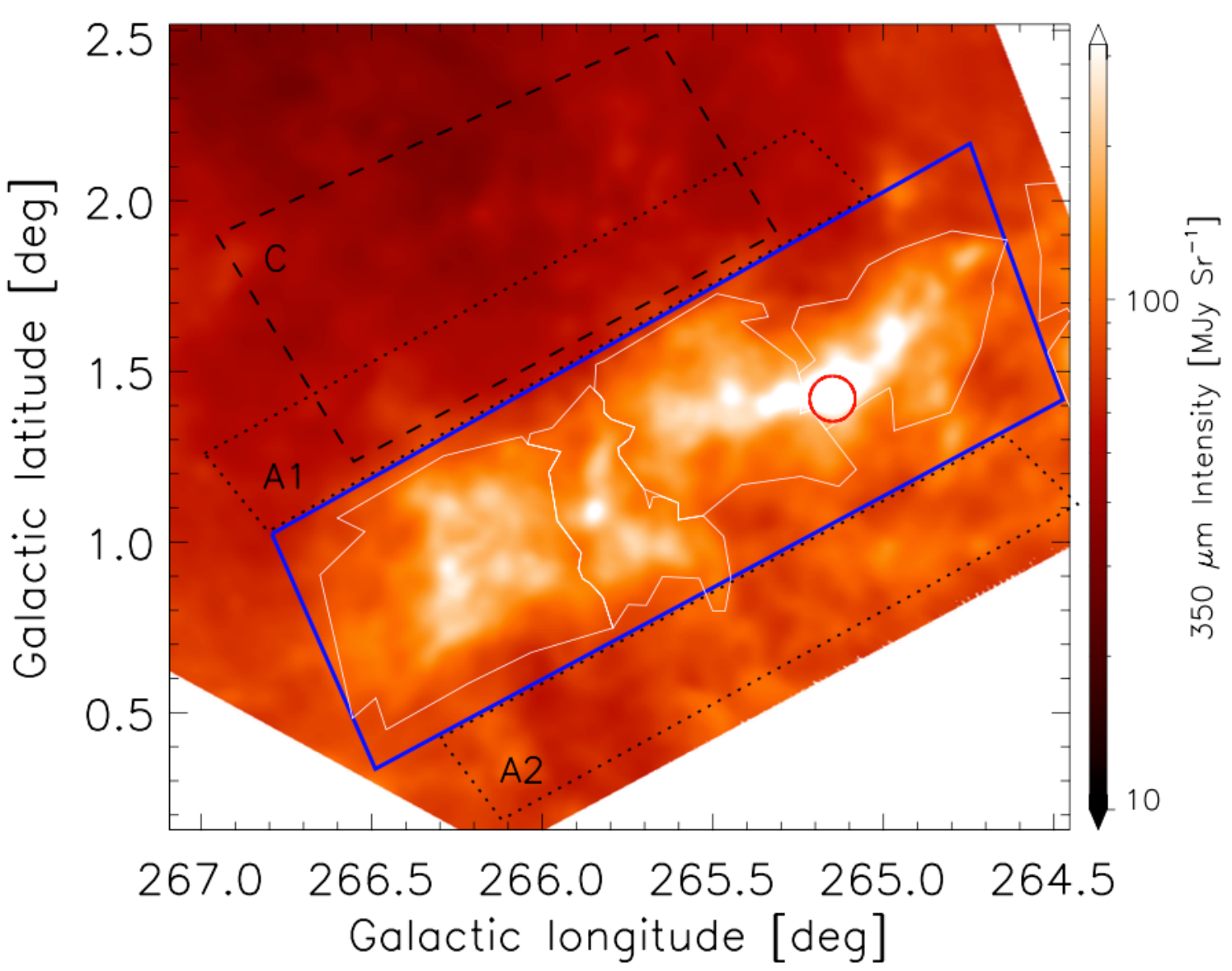}
\caption{\blastpol~350\,\um~intensity map of Vela C showing various regions used in the analysis (see the text). \label{regions}}
\end{figure}

Polarimetry data at 250, 350, and 500\,\um~were collected during the 2012 flight of \blastpol~\citep{Galitzki2014,Fissel2015}. \blastpol~was launched on a stratospheric balloon on 2012 December 26 from McMurdo Station, Antarctica and took data for 12.5 days at an altitude of 38.5 km above sea level. Approximately 43 hr were spent mapping the Vela C molecular cloud (Figure \ref{regions}), covering four of the five sub-regions defined by \citet{Hill2011}. The data analysis pipeline and instrument characterization are described in \citet{Fissel2015}.

The \Planck HFI instrument \citep{PlanckI} obtained polarimetry data at 850\,\um ~(their 353 GHz band) over the whole sky \citep{PlanckXIX}. Data from the available map on the Planck Legacy Archive \footnote{http://pla.esac.esa.int} were regridded to match the \blastpol~maps. To match the resolution of the \Planck map, the \blastpol~maps were smoothed to $4\farcm8$. Maps were sampled in pixels of size $2\farcm4$.

To focus the analysis on dust within the Vela C molecular cloud itself, it is necessary to subtract the emission (in $I$, $Q$, and $U$) from dust along the line of sight and in the extended Vela Molecular Ridge.
We adopted two subtraction methods described in \citet{Fissel2015} and their terminology: ``conservative'' and ``aggressive,'' and a third, ``intermediate,'' for the average of the two.  The first uses a single reference region, labelled C in Figure~\ref{regions}, while the second interpolates a plane between two regions labelled A1 and A2 bracketing the outlined ``validity region."

The data were restricted to being inside the Vela C sub-regions defined in \citet{Hill2011}, because these sightlines are better probes of the polarization structure in the cloud itself. These are shown in white in Figure~\ref{regions}. Data outside these regions are also more sensitive to systematic uncertainties in the method of diffuse emission subtraction. Data from a circular region, shown as a red circle in Figure~\ref{regions}, within $4\arcm$ of RCW 36 ($l=265\fdg15, b=1\fdg42$) were also excluded because null tests showed large residuals there.

\begin{figure}
\epsscale{1.2}
\plotone{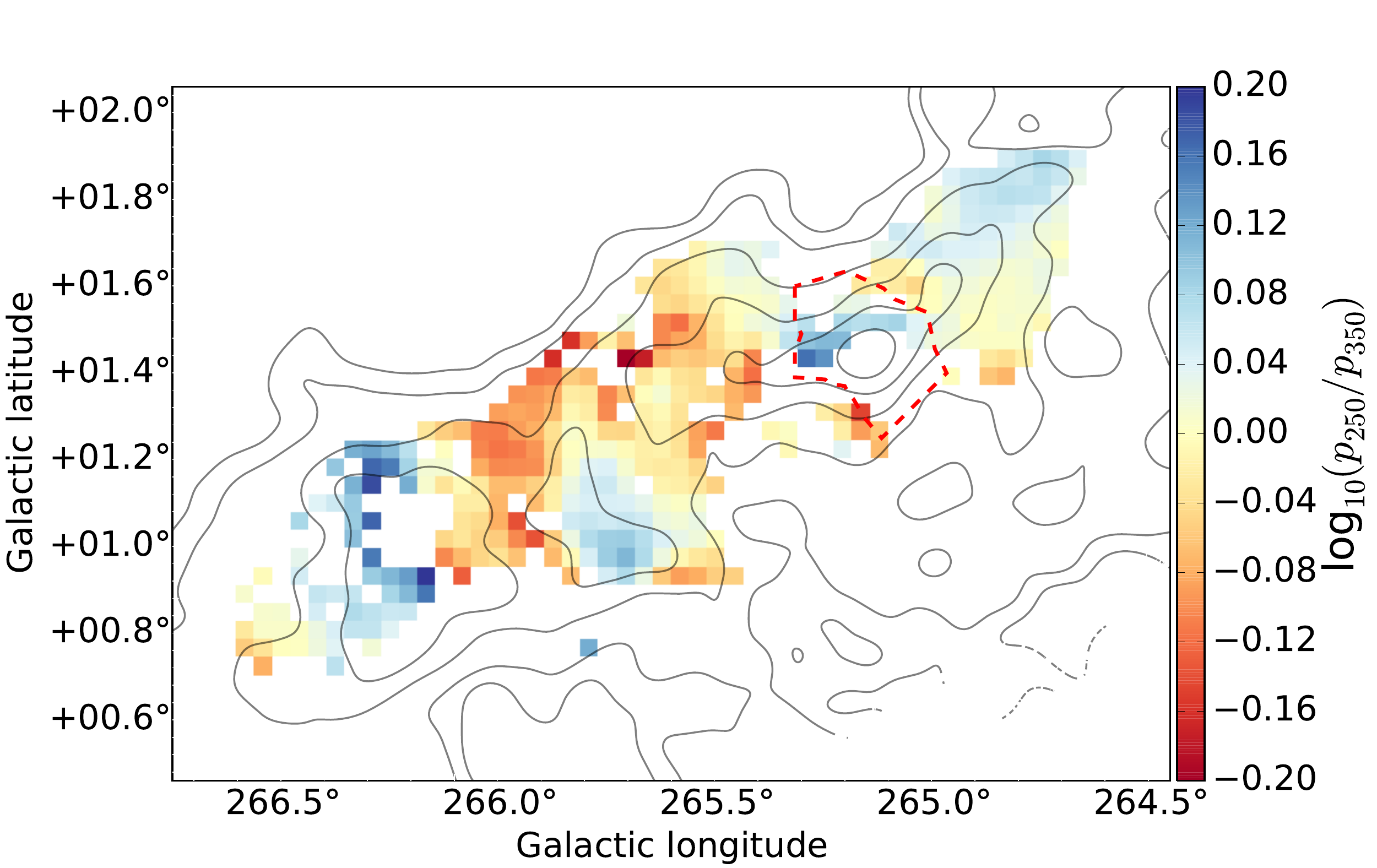}
\plotone{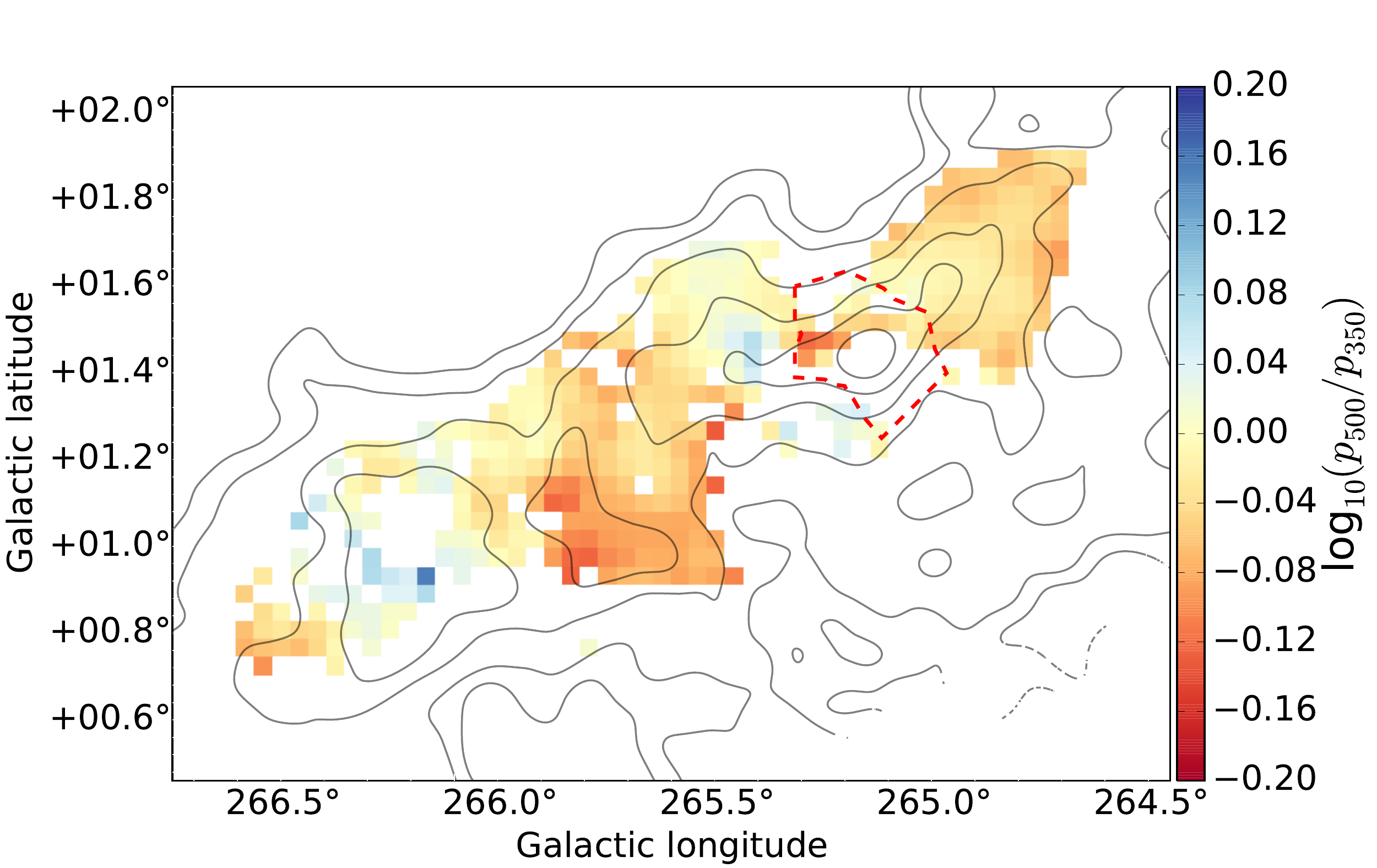}
\plotone{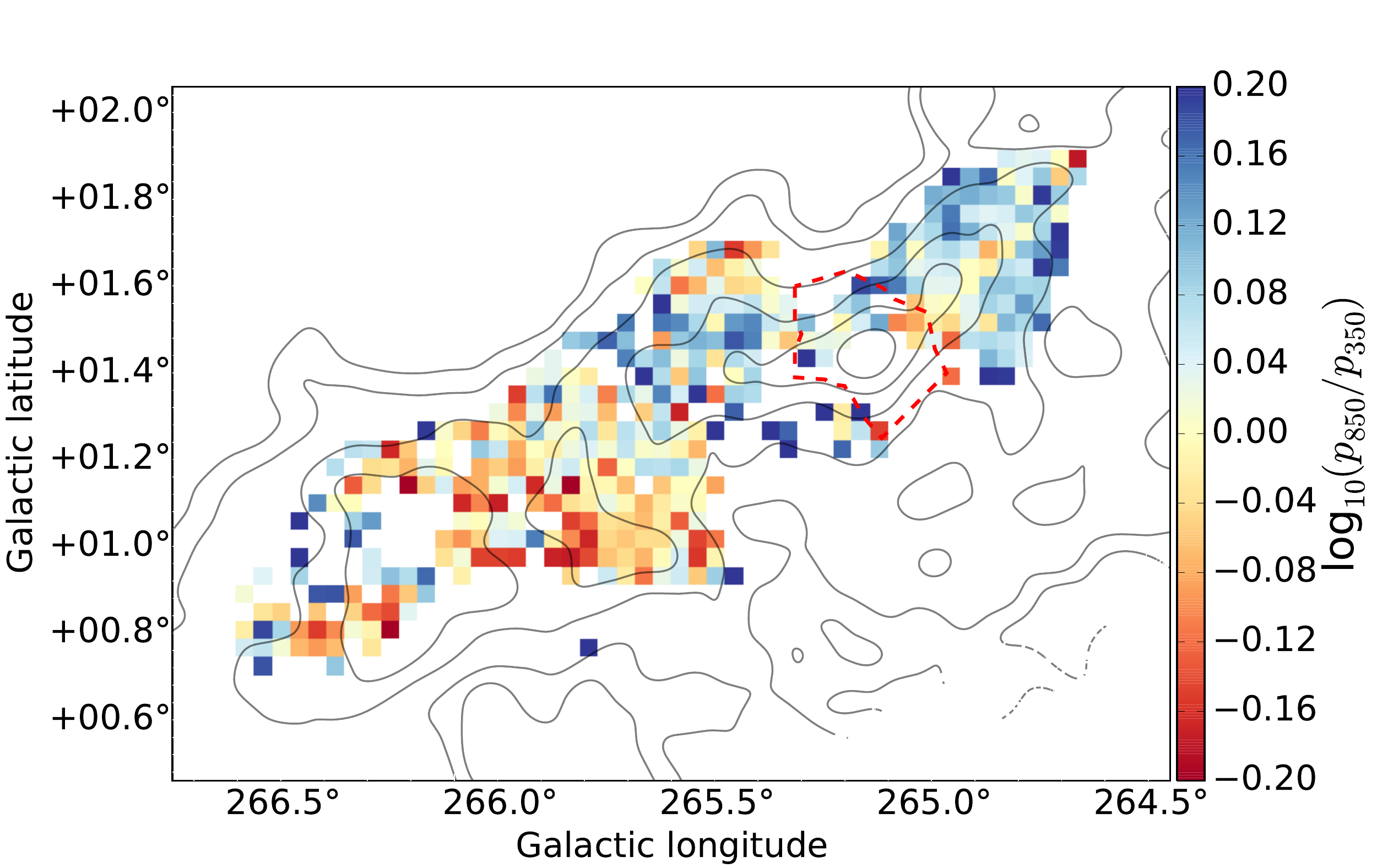}
\caption{Maps of polarization ratios \pratone~(top), \prattwo~(middle) and \pratthree~(bottom). The color bar is on a logarithmic scale. Contours show 350\,\um~intensity at levels of 1\%, 2\%, 5\%, 10\%, 20\%, and 50\% of the peak intensity. The red dashed line indicates the region used to isolate the dust being irrradiated by RCW 36 (see Section \ref{sec:envt}).\label{pratios}}
\end{figure}

The polarization fraction, $p$, was calculated at 250, 350, 500, and 850\,\um~from the Stokes parameters $I$, $Q$, and $U$ as $p=\sqrt{Q^2+U^2}/I$. Because the polarization amplitude is a positive definite quantity, noise in the $Q$ and $U$ maps will tend to increase the measured polarization. The measured value, $p_{\mbox{\footnotesize m}}$, was therefore approximately corrected using \citep{Wardle1974}:
\begin{equation}
p_{\mbox{\footnotesize db}}\,=\,\sqrt{p_{\mbox{\footnotesize m}}^2\,-\,\sigma_{\mbox{\footnotesize p}}^2}\,,
\end{equation}
where $p_{\mbox{\footnotesize db}}$ is the de-biased polarization and $\sigma_{\mbox{\footnotesize p}}$ is the uncertainty associated with $p_{\mbox{\footnotesize m}}$. This is a crude approximation, but reasonably accurate for high signal-to-noise ratio measurements \citep{Montier2015}, such as the data analyzed in this work where only $p_{\mbox{\footnotesize m}}\geq3\sigma_{\mbox{\footnotesize p}}$ was used. 

The polarization angle, $\psi$, with respect to the $Q,U$ reference frame, was determined as $\psi\,=\,\frac{1}{2} \arctan\left(U,Q\right)$. The polarization spectrum was studied only where the measurements of $p$ exceeded $3\sigma_{\mbox{\footnotesize p}}$ at each of the four wavelengths, and the measured polarization angles at all wavelengths agreed to within 1$0\deg$. This last cut has been used in previous studies of submillimeter polarization \citep{Vaillancourt2012}, and reduces the probability that data at the different wavelengths are measuring different points along the line of sight.

The number of data points passing the $p\geq3\sigma_{\mbox{\footnotesize p}}$ cut depends on the method of diffuse emission subtraction used; however, for all of the subtraction methods none of the remaining data fail the $10\deg$ angle cut. In total the number of data points included in the analysis is 383, 405, and 403 in the case of conservative, aggressive, and intermediate subtraction, respectively, out of a total of 615 measurements lying inside the regions described earlier in this section.

Ratios of $p$ were then calculated relative to $p_{350}$, in order to compare with previous studies which also normalized their measurements to 350\,\um. Figure \ref{pratios} shows maps of the polarization ratios \pratone, \prattwo, and \pratthree~for the case where the intermediate subtraction method was applied. The 850\,\um~data from \Planck have a higher intrinsic noise, resulting in a higher noise level in the \pratthree~map.

\begin{figure}
\begin{centering}
\epsscale{1}
\plotone{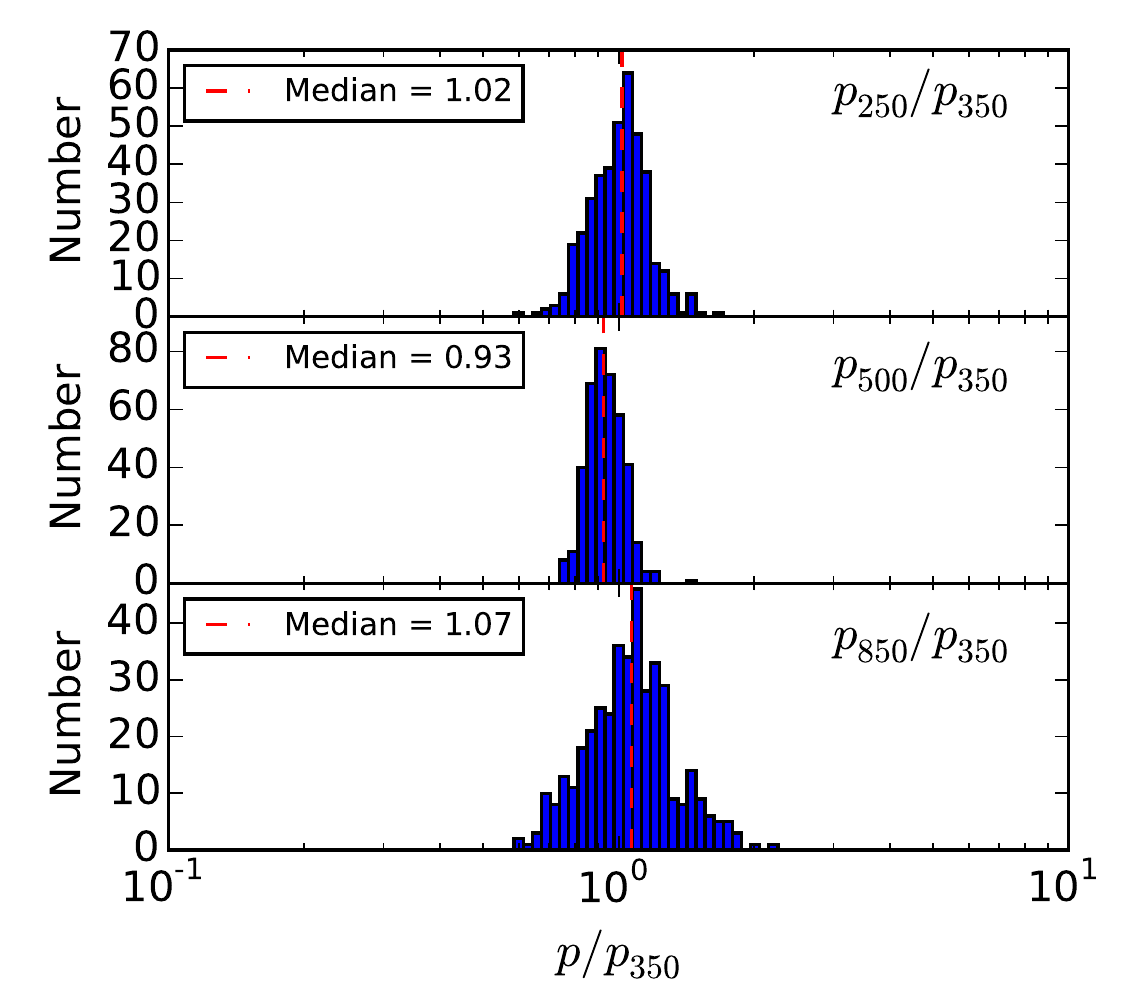}
\caption{Histograms of the three polarization ratios, using the intermediate diffuse emission subtraction method. Median polarization ratios are indicated by dashed lines.\label{medhistos}}
\end{centering}
\end{figure}

\begin{figure}
\epsscale{1.25}
\plotone{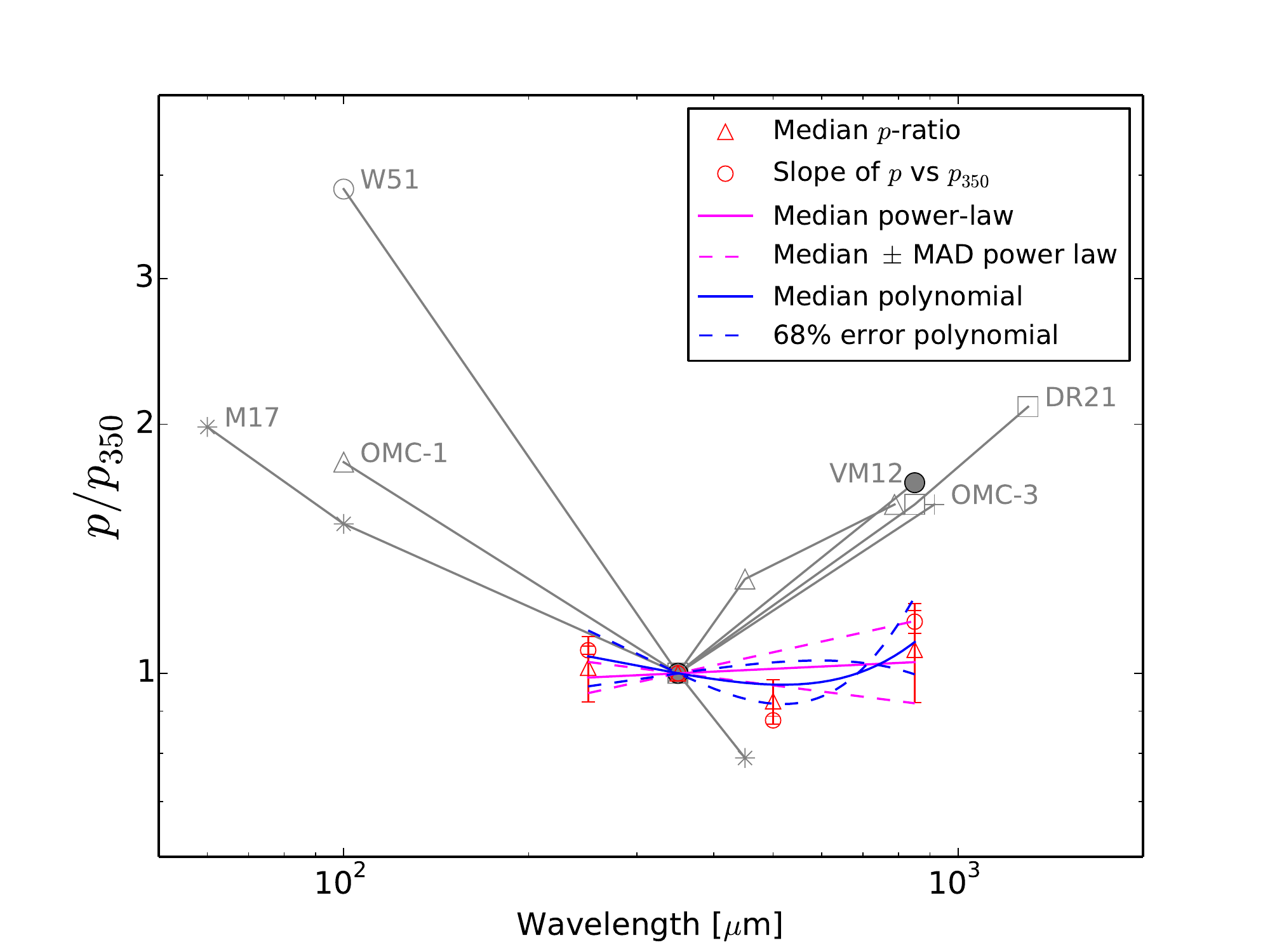}
\caption{Polarization spectra from previous work (gray), with new Vela C data added (colors). Points at 850\,\um~separated horizontally for clarity. W51, OMC-1 $p_{100}/p_{350}$, and DR21 $p_{1300}/p_{350}$ from \citet{Vaillancourt2002}. All previous measurements of $p_{850}/p_{350}$ from \citet{Vaillancourt2012}. The solid circle represents their median ratio for 15 clouds. OMC-1 $p_{450}/p_{350}$ from \citet{Vaillancourt2008}. M17 from \citet{Zeng2013}. Red triangles are median polarization ratios with MAD error bars. Red circles are best-fit slopes to scatter plots of $p/p_{350}$. Magenta lines are spectra using the power-law fit parameters, and blue lines are spectra using the second-order polynomial fit parameters. Solid lines use the median values of the fit parameters and dashed lines reflect the distribution in the fit parameters (see the text). \label{allspec}}
\end{figure}

\section{Results}
\label{sec:results}
\subsection{Median polarization ratios}
\label{sec:medians}

Figure \ref{medhistos} shows histograms of the three polarization ratios $p_{250}/p_{350}$, $p_{500}/p_{350}$, and $p_{850}/p_{350}$ for the case of intermediate diffuse subtraction, with the median values indicated. As in previous submillimeter polarization spectrum studies mentioned in Section \ref{sec:Intro}, the distributions of polarization ratios are non-normal, with several outliers. Therefore, the central tendencies of the histograms are better represented by the median values than by the means. The median absolute deviation was used to quantify the scatter in the distribution:
\begin{equation}
\mbox{MAD}\equiv\mbox{median}(|x-x_{\mbox{\footnotesize m}}|)\,,
\end{equation}
where $x_{\mbox{\footnotesize m}}$ is the median value of the measurements $x$. Table \ref{table:medians} lists the medians and MADs for the three polarization ratios using the three diffuse emission subtraction methods. For the intermediate subtraction method, the ratios were \pratone~= 1.02$\pm$0.09, \prattwo~= 0.93$\pm$0.06, and \pratthree~= 1.07$\pm$0.15. Although there is a slight minimum at 500\,\um, the median $\pm$ MAD polarization ratios at all of the wavelengths observed are not significantly different from each other, and they are all very close to a flat spectrum, i.e. a ratio of 1.0, as can be visualized in Figure \ref{allspec} (red triangles). This result does not depend on the method of diffuse emission subtraction.

\begin{table}
\begin{center}
\caption{Medians and MADs of polarization ratios ($p_\lambda/p_{350}$)}
\begin{tabular}{@{}lccc@{}}
\toprule
Diffuse Emission&250\,\um&500\,\um&850\,\um\\
Subtraction Method&&&\\
\midrule
Conservative&$0.97\pm0.12$&$0.93\pm0.05$&$1.04\pm0.14$\\
Aggressive&$1.07\pm0.08$&$0.92\pm0.06$&$1.08\pm0.16$\\
Intermediate&$1.02\pm0.09$&$0.93\pm0.06$&$1.07\pm0.15$\\
\bottomrule
\end{tabular}
\end{center}
\label{table:medians}
\end{table}

\subsection{Polarization ratios from scatter plots}\label{sec:slopes}
\begin{figure}
\epsscale{1}
\plotone{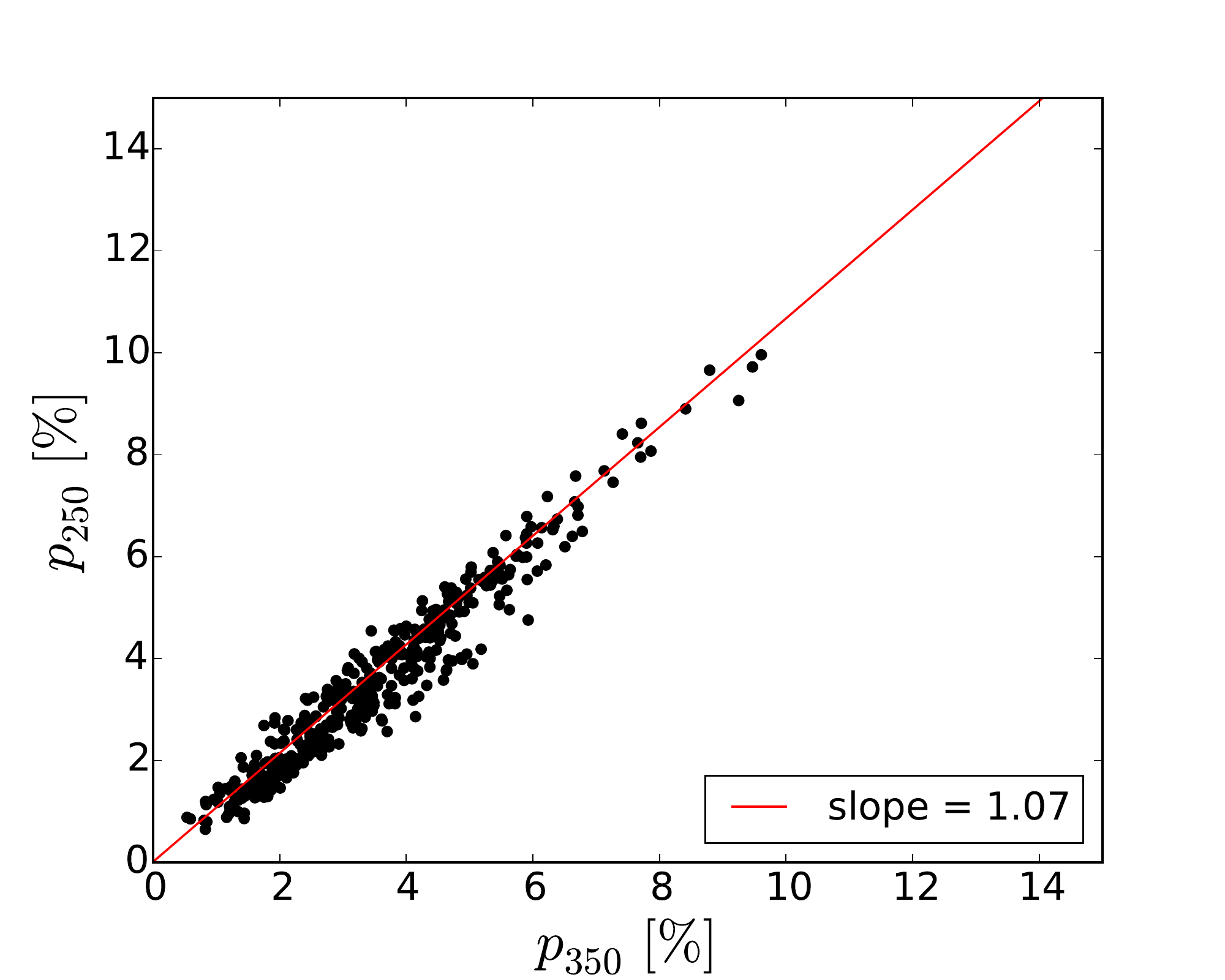}
\plotone{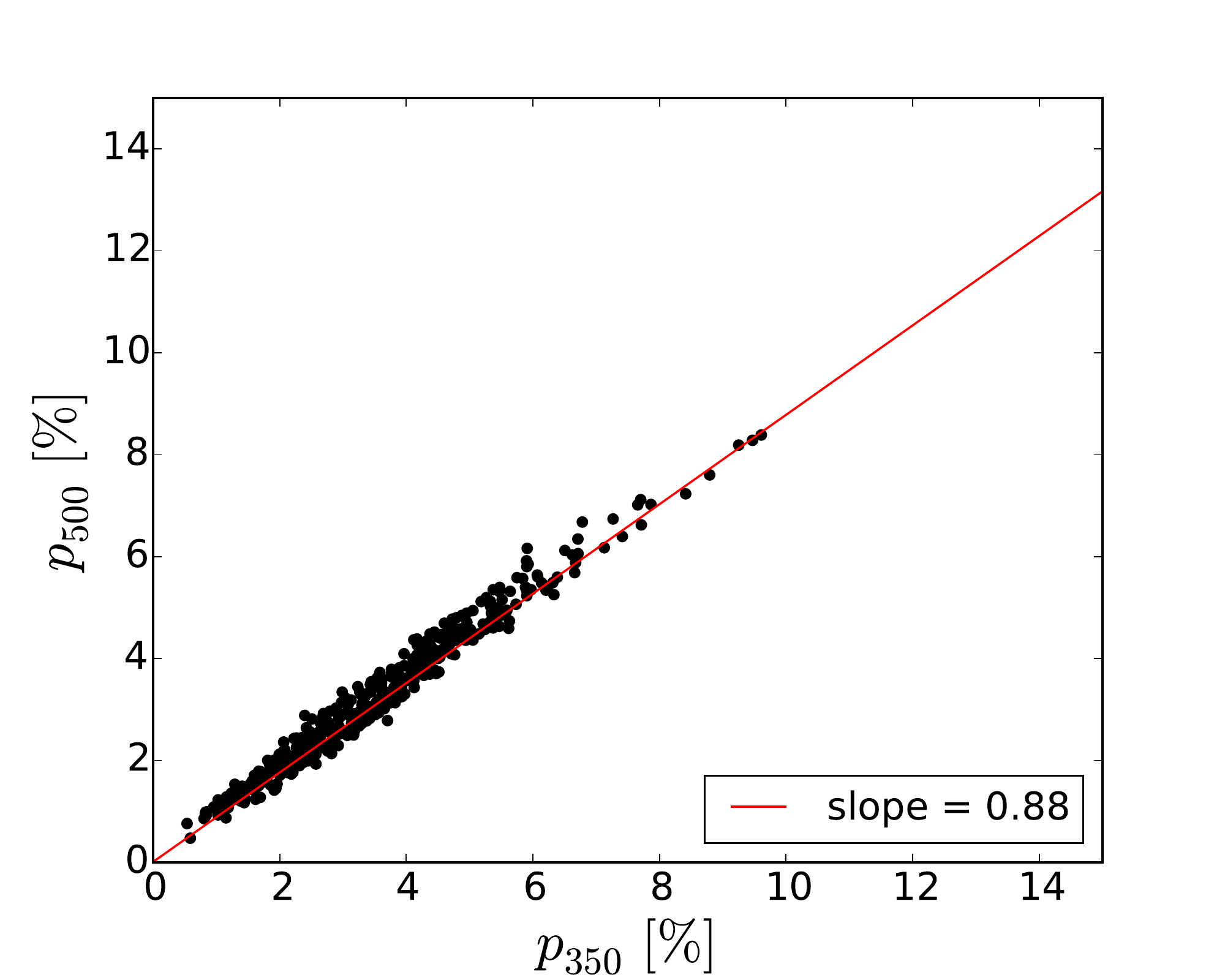}
\plotone{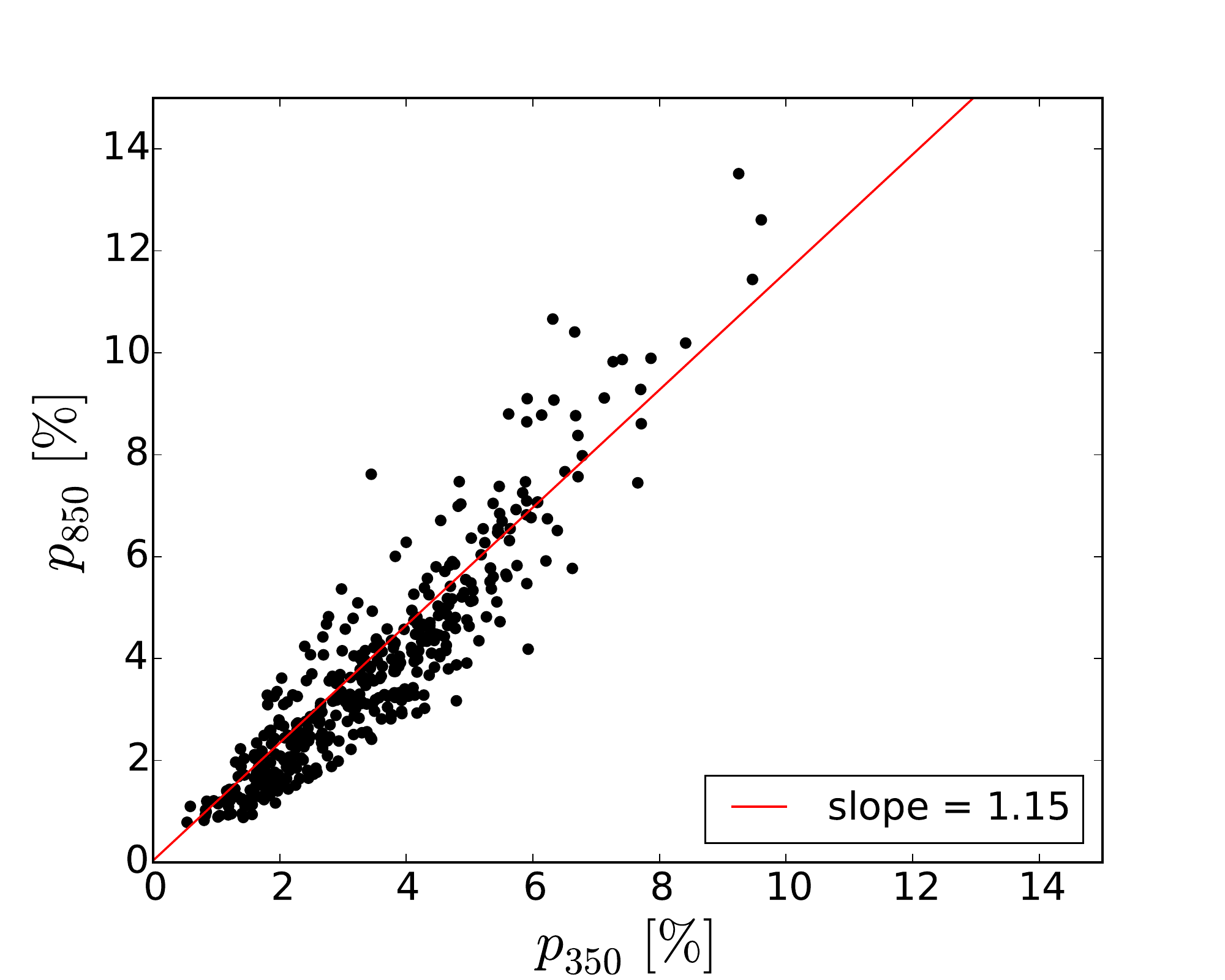}
\caption{Linear fits to scatter plots of $p_\lambda$ vs $p_{350}$. The red line indicates the best fit to a linear model with the slope indicated.\label{scatterplots}}
\end{figure}

As an alternative, linear fits were performed on three scatter plots of polarization fractions: 250\,\um~versus 350\,\um, 500\,\um~versus 350\,\um, and 850\,\um~versus 350\,\um~(Figure \ref{scatterplots}). For each, the data were fit by minimizing the absolute deviation between the data and a linear model $p_{\lambda}=a_{\lambda}+b_{\lambda}~p_{350}$. The slopes, $b_{\lambda}$, of each line constitute a polarization spectrum.

The least absolute deviation was used rather than a least squares method because it produces a more robust solution, due to its resistance to outliers. The uncertainty on the slopes was calculated using the bootstrapping method with replacement \citep{nr_boot}, repeating the fits for each of 10,000 random selections. The standard deviation of the distribution of slopes is used as an estimate of the uncertainty.

Table \ref{table:slopes} lists the slopes and their uncertainties for each of the three scatter plots and for each of the three diffuse emission subtraction methods. The spectrum has a minimum at 500\,\um~(see Figure \ref{allspec}, red circles), independent of the method of diffuse emission subtraction. Again, the spectrum is flat to within about 10\%.

Comparing the values in Tables \ref{table:medians} and \ref{table:slopes} shows that the slopes obtained by fitting to scatter plots are consistent with the medians $\pm$ MADs obtained in the previous section. However, the uncertainties on the slopes in Table \ref{table:slopes} are small compared to the values cited in Table \ref{table:medians} for the MADs. These are quite different approaches, here describing the uncertainty in the linear fit to the data, in contrast to the MAD representing the width of the distribution of the ratioed data. It is not surprising that the latter are much larger. In the analogous situation in which the errors were normally distributed, one would calculate the mean and standard deviation about the mean, and then estimate the uncertainty of the mean by dividing the standard deviation by the square root of the number of data.

\begin{table}
\begin{center}
\caption{Slopes of linear fits to scatter plots of $p_\lambda$ vs $p_{350}$}
\begin{tabular}{@{}lccc@{}}
\toprule
Diffuse Emission&250\,\um&500\,\um&850\,\um\\
Subtraction Method&&&\\
\midrule
Conservative&$1.12\pm0.01$&$0.89\pm0.01$&$1.15\pm0.04$\\
Aggressive&$1.04\pm0.01$&$0.87\pm0.01$&$1.19\pm0.04$\\
Intermediate&$1.07\pm0.01$&$0.88\pm0.01$&$1.15\pm0.04$\\
\bottomrule
\end{tabular}
\end{center}
\label{table:slopes}
\end{table}

\begin{table*}
\begin{center}
\caption{Median and MADs of $p(\lambda)$ fit parameters}
\begin{tabular}{@{}lccccccc@{}}
\toprule
Diffuse Emission &Linear Fit& &Power-law Fit & &Polynomial Fit & & \\
Subtraction Method &$b_l (\times10^{-4})$& $p_{350}/a_l$ & $b_{pl}$ & $p_{350}/a_{pl}$ &$b_{2p} (\times10^{-6})$&$c_{2p} (\times10^{-4})$ & $p_{350}/a_{2p}$\\
\midrule
Conservative&\conslinearb&\conslineara&\conspowerlawb&\conspowerlawa&\conspolyb&\conspolyc&\conspolya \\
Aggressive&\agglinearb&\agglineara&\aggpowerlawb&\aggpowerlawa&\aggpolyb&\aggpolyc&\aggpolya \\
Intermediate&\intlinearb&\intlineara&\intpowerlawb&\intpowerlawa&\intpolyb&\intpolyc&\intpolya \\
Intermediate $(p/p_{350})$&\cloudlinearb&\dots&\cloudpowerlawb&\dots&\cloudpolyb&\cloudpolyc&\dots \\
\bottomrule
\end{tabular}
\end{center}
\label{table:fits}
\end{table*}

\subsection{Fits to $p(\lambda)$}\label{sec:lfits}
To explore the shape of the spectrum further, fits were performed to $p(\lambda)$ at each individual pixel in the map using the measurements of $p$ at 250, 350, 500, and 850\,\um. The data were fit to three different functions of $p(\lambda)$; linear
\begin{equation}\label{eq:linfit}
p(\lambda)=a_l[b_l(\lambda-\lambda_0)+1]\,;
\end{equation}
power law
\begin{equation}\label{eq:powlawfit}
p(\lambda)=a_{pl}\left(\frac{\lambda}{\lambda_0}\right)^{b_{pl}}\,;
\end{equation}
and a second-order polynomial
\begin{equation}\label{eq:polyfit}
p(\lambda)=a_{2p}[b_{2p}(\lambda-\lambda_0)^2+c_{2p}(\lambda-\lambda_0)+1]\,.
\end{equation}
Here $\lambda_0=350$\,\um~and in each case $a$ is a normalization constant that is factored out. The linear and power-law fits are different attempts to measure the overall increase or overall decrease of the spectrum in the 250-850\,\um~range. However, in addition, the second-order polynomial fit also allows in addition a spectrum that has a minimum or maximum between 250 and 850\,\um. Although the error bars for $p$ at 250, 350, and 500\,\um~are much smaller than at 850\,\um, the overall uncertainty is dominated by the diffuse emission subtraction method which affects all four bands. Therefore, each of the bands was given equal weight in the fits to $p(\lambda)$.

\begin{figure}
\epsscale{1.0}
\plotone{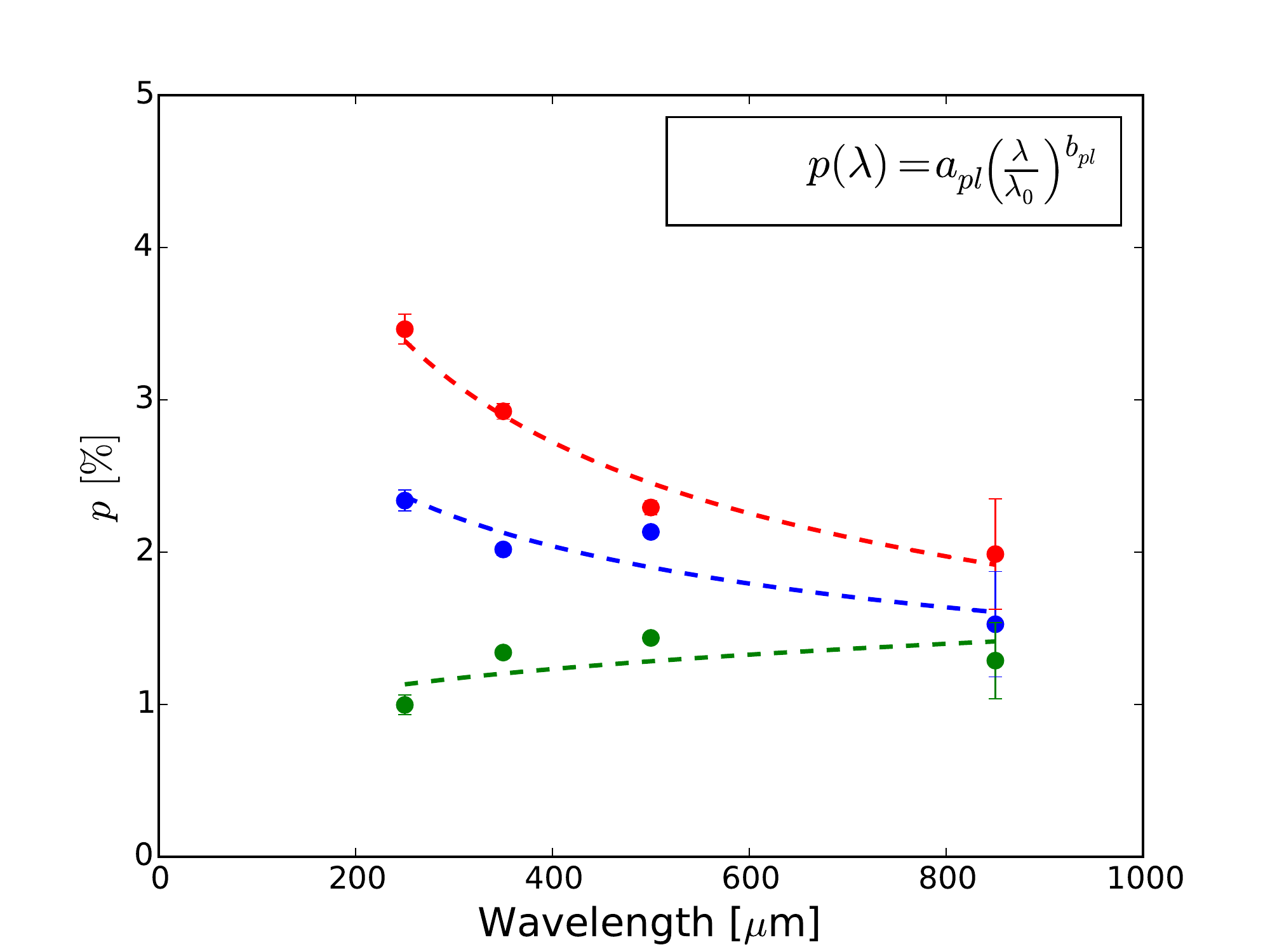}
\plotone{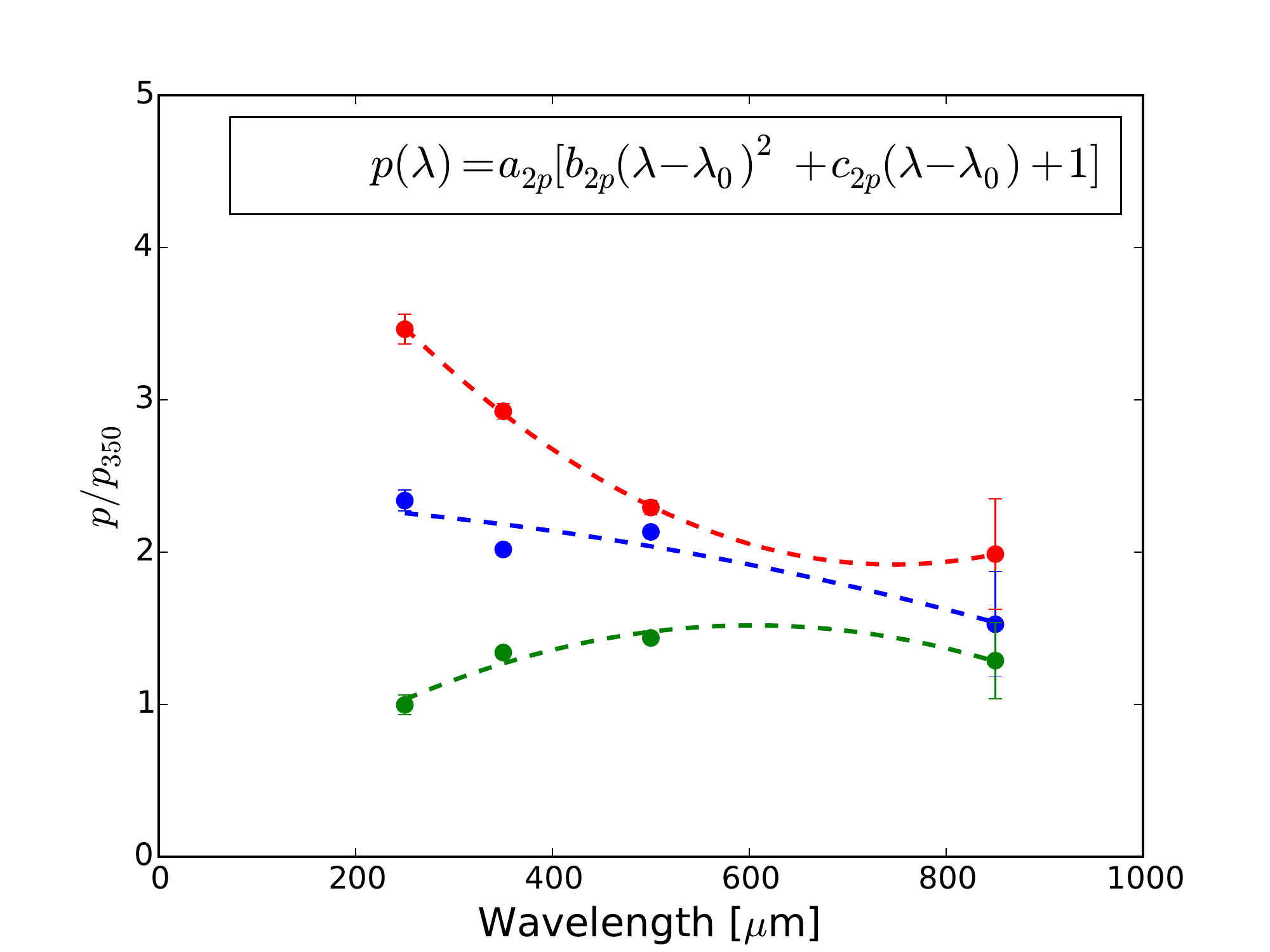}
\caption{Example fits of $p(\lambda)$ to the data from three pixels at $l = 265\fdg79, b = 1\fdg01$ (red), $l = 266\fdg29, b = 0\fdg85$ (blue), and $l = 266\fdg08, b = 0\fdg93$ (green). Top: power-law fit (Equation~(\ref{eq:powlawfit})). Bottom: second-order polynomial fit (Equation~(\ref{eq:polyfit})).\label{plambdafits}}
\end{figure}

Figure \ref{plambdafits} shows the results of fitting the power-law and second-order polynomial functions to the data for three example pixels. The linear fit is not shown, because the linear and power-law spectrum look very similar in the 250-850\,\um~range. 

\begin{figure}
\epsscale{1}
\plotone{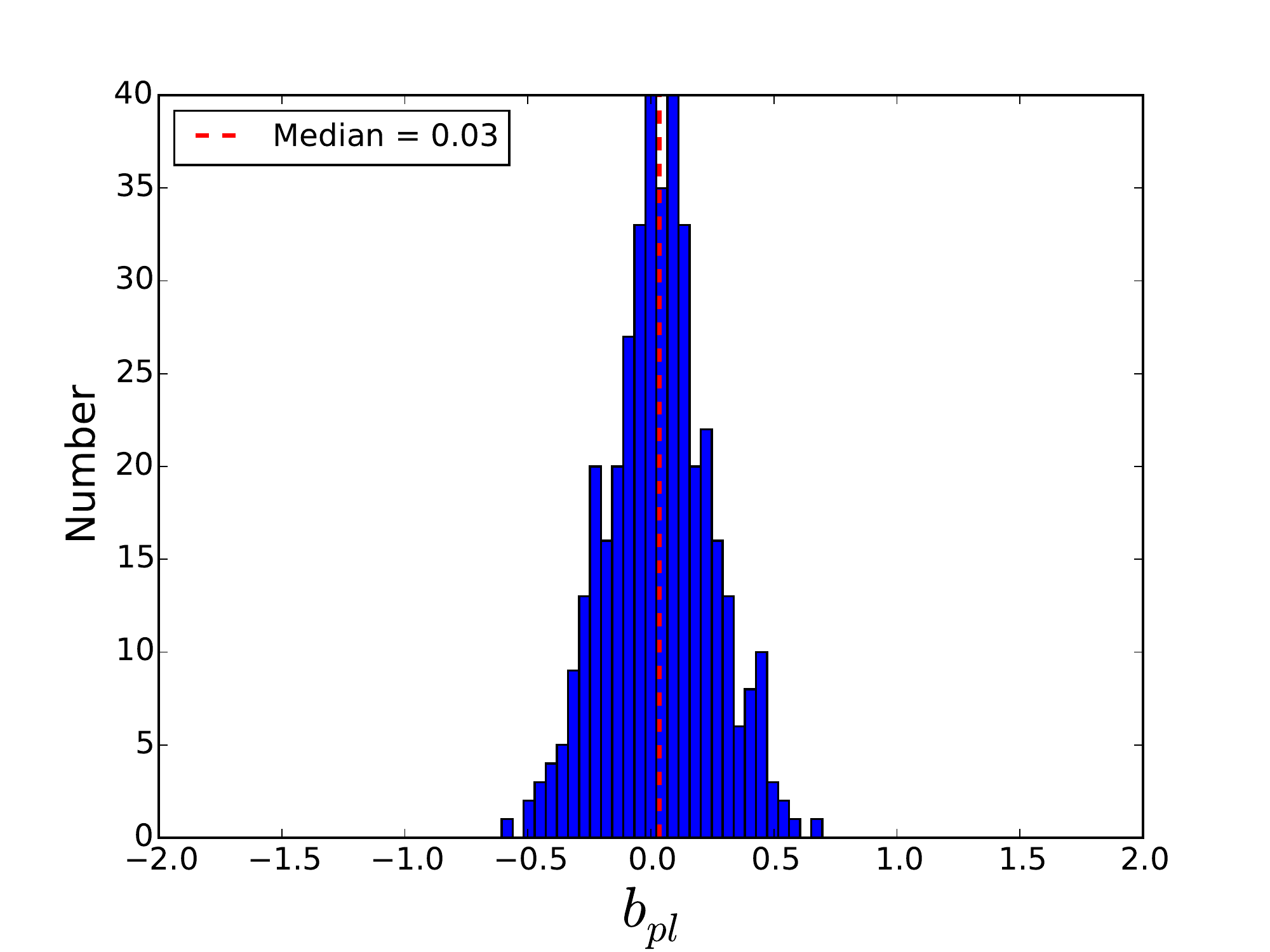}
\caption{Distribution of power-law exponent $b_{pl}$ from Equation~(\ref{eq:powlawfit}).\label{powerparams}}
\end{figure}

\begin{figure}
\epsscale{1}
\plotone{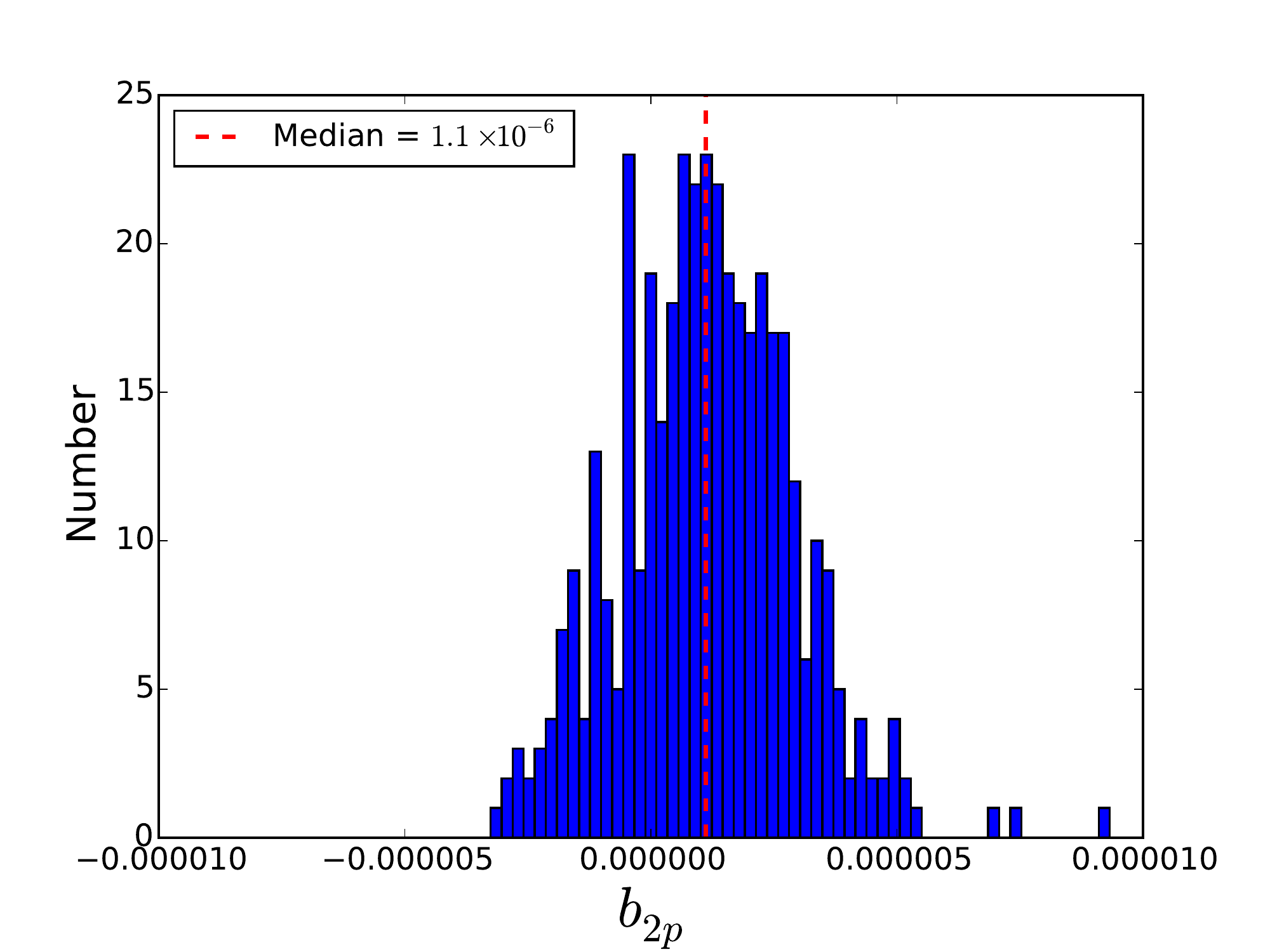}
\plotone{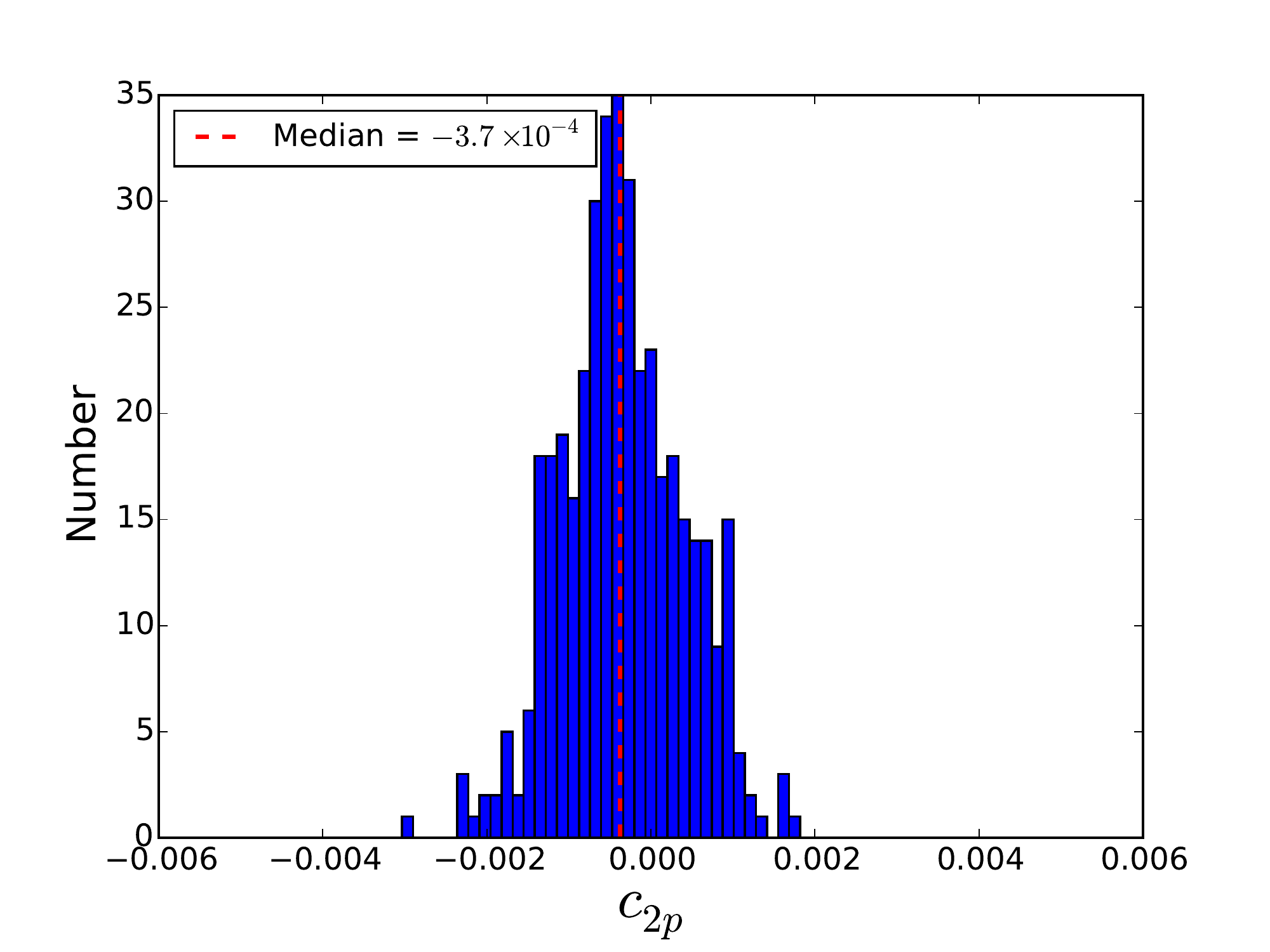}
\caption{Distributions of polynomial fit parameters $b_{2p}$ (top) and $c_{2p}$ (bottom) from Equation~(\ref{eq:polyfit}). \label{polyparams}}
\end{figure}

The distribution of the fit parameters over all pixels in the Vela C map was then analyzed. The first three rows of Table \ref{table:fits} list the median values obtained for each of the fit parameters that relate to the spectrum shape ($b_l$, $b_{pl}$, $b_{2p}$, and $c_{2p}$), for the three methods of diffuse emission subtraction. Table \ref{table:fits} also contains the median values of $p_{350}/a$, showing how closely the fits match the data at 350\,\um.  The typical fractional uncertainty of the measurement of $p_{350}$ is compatible with the MAD. The values of the spectral shape fit parameters are consistent among the three subtraction methods, and their distributions for the case of intermediate subtraction are shown in Figures \ref{powerparams} (power-law fit) and \ref{polyparams} (polynomial fit).

The median of the linear fit parameters over the cloud produces a spectrum that gradually rises from \pratone~= 0.99 to \pratthree~= 1.03. The median power-law fit produces a spectrum that is virtually identical to the median linear fit. The second-order polynomial fit was used to probe curvature in the spectrum, looking for a clear minimum or maximum. The median polynomial fit produces a spectrum with a minimum of $p/p_{350}\sim0.97$ at $\lambda\sim518$\,\um, rising to \pratone~$\sim1.05$ and \pratthree~$\sim1.10$. 

Figure \ref{allspec} shows polarization spectra over the 250 to 850\,\um~range plotted using the median $p(\lambda)$ power-law and second-order polynomial fit parameters for the intermediate subtraction method. The dispersion among these fits is illustrated by constructing spectra using the median~$\pm$~MAD values of the fit parameters. For the polynomial fit, a scatter plot of $b_{2p}$ vs. $c_{2p}$ shows that the two terms are highly anti-correlated. A 68\% error ellipse was fit to the distribution; the two endpoints of the major axis of the ellipse were $(b_{2p},c_{2p})=(-6.1\times10^{-7},2.9\times10^{-4})$ and $(b_{2p},c_{2p})=(2.9\times10^{-6},1.0\times10^{-3})$. These values were used to construct the dashed blue curves shown in Figure \ref{allspec}. We note that the curves plotted all pass through 1.0 by construction and, from the values of $p_{350}/a$, that the curves typically fit the data to within 4\,\% at 350\,\um, which is about the size of the symbols plotted there.

An alternative method is to fit to the slopes $p/p_{350}$ from Table \ref{table:slopes}, using their associated uncertainties as weights in the fit. Equations (\ref{eq:linfit})-(\ref{eq:polyfit}) were modified to the normalized form with no ``$a$" fitting parameter and there was no 350\,\um~data point. The remaining fit parameters obtained using this method are listed at the bottom of Table \ref{table:fits}. The first two types of fits to $p$ have a negative slopes reflecting the down-weighting of the 850\,\um~data. However, the polynomial fit is not greatly changed, with a minimum of $p/p_{350}\sim0.92$ at $\lambda\sim535$\,\um, rising to \pratone~$\sim1.10$ and \pratthree~$\sim1.15$.

\subsection{Effect of Environment}\label{sec:envt}
The environment in the Vela C molecular cloud can be represented quantitatively by column density of hydrogen nuclei, $N\,[\mbox{cm}^{-2}]$ and dust temperature, $T\,[\mbox{K}]$. These quantities were derived from modified blackbody SED fits to \Herschel data at 160, 250, 350, and 500\,\um, assuming a dust spectral index $\beta=2$. The methodology is described in detail in \citet{Fissel2015}. For this analysis, the \Herschel maps were smoothed to $5\arcm$ before the SED fitting.

\begin{figure*}
\epsscale{1.0}
\plotone{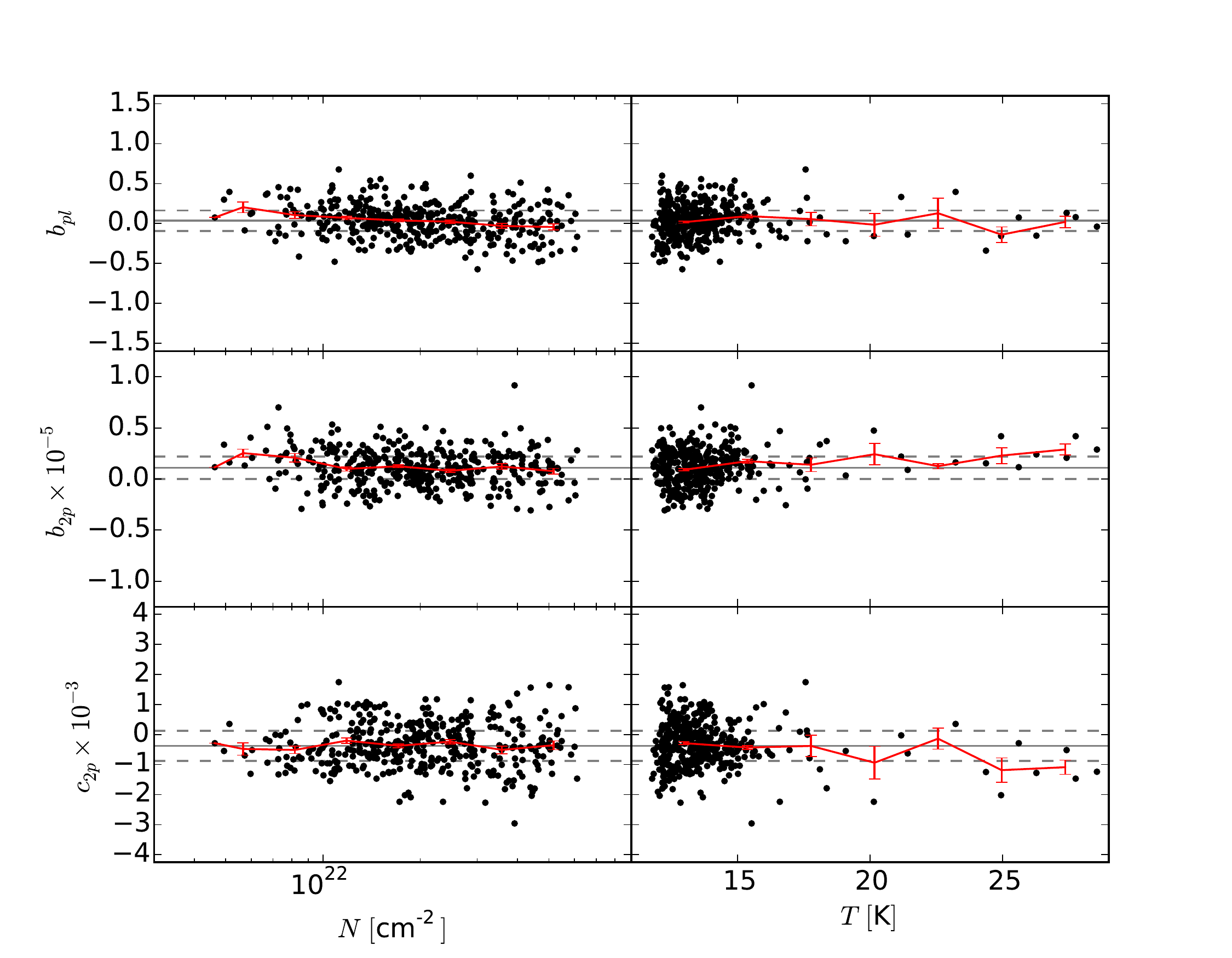}
\caption{Fit parameters versus column density (left) and temperature (right). From top to bottom: power-law index $b_{pl}$; second-order polynomial coefficient $b_{2p}$; and first-order polynomial coefficient $c_{2p}$. The gray lines show the median $\pm$ MAD values of the fit parameters. Red lines join means of the fit parameters in bins spaced logarithmically in $N$ and linearly in $T$. Error bars indicate the standard deviation of each bin.\label{plambdavsNT}}
\end{figure*}

To investigate whether the shape of the polarization spectrum depends on environment, the $p(\lambda)$ fit parameters obtained in Section \ref{sec:lfits} were plotted versus $N$ and $T$. The results are shown in Figure \ref{plambdavsNT} for the power-law and polynomial fit parameters. The data were binned in $N$ and $T$ and the mean value of each parameter plotted for each bin. No trend is seen with $N$, i.e. for all of the $N$ bins, the average value stays within the median $\pm$ MAD listed in Table \ref{table:fits}. No trend is seen over most of the $T$ bins, but for the sparse data at high $T$ the polynomial fit gives a slightly higher second-order coefficient, with a correspondingly lower value of the anti-correlated first-order coefficient.

The high $T$ data come from the region of the map around the compact \ion{H}{2} region RCW 36, and so the effect of environment was also examined by separating the cloud into two regions. In one region, the dust is heated by RCW 36, and in the other it is heated by the interstellar radiation field (ISRF). \citet{Fissel2015} describe how the separation of the two populations was done, based on the observation of a clear trend of decreasing $T$ with increasing $N$ for dust heated by the ISRF. Data that do not lie along this trend line are from sightlines heated by RCW 36. 
The closed curve that separates the two regions is shown in Figure \ref{pratios}. The numbers of data points were 25 (RCW 36) and 434 (ISRF). The same methods were then applied to each of the two regions; the results are listed in Table \ref{table:2pops}. 

\begin{figure}
\epsscale{1.25}
\plotone{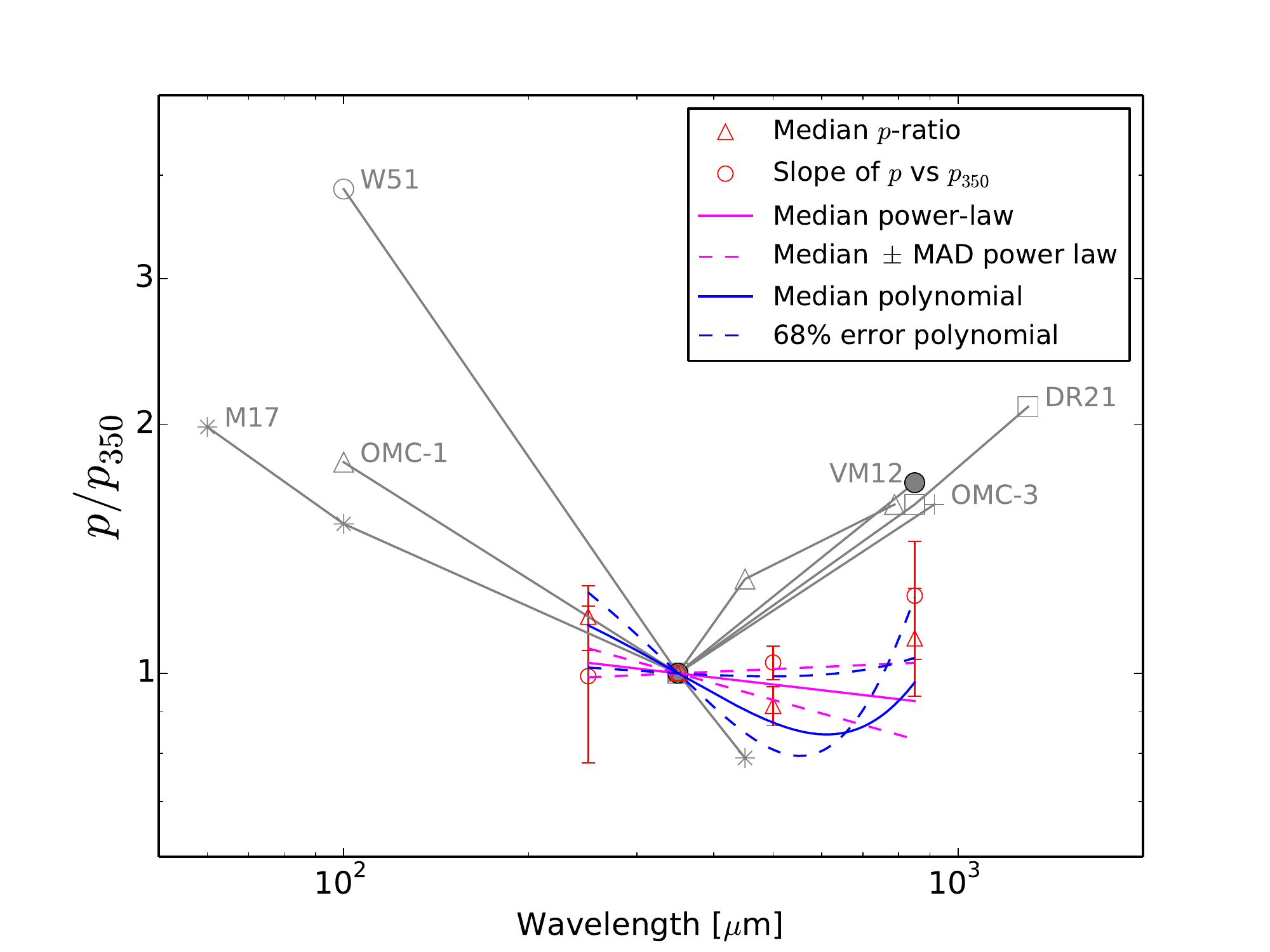}
\caption{Polarization spectra for RCW 36-heated sight lines in Vela C. See caption to Figure \ref{allspec} for explanation of symbols used.\label{2pops}}
\end{figure}

The results for the ISRF-heated sightlines are virtually identical to those for the Vela C cloud as a whole. The results for the two groups are also consistent with each other within their uncertainties. Figure \ref{2pops} shows the spectrum for the RCW 36-heated sightlines. 

\begin{table}
\begin{center}
\caption{Results for sightlines heated by RCW 36 vs sightlines heated by the ISRF}
\begin{tabular}{lcc}
\toprule
Measurement Quantity & RCW 36 & ISRF\\
\midrule
Median \pratone &$1.17\pm0.11$&$1.01\pm0.09$\\
Median \prattwo &$0.91\pm0.05$&$0.93\pm0.06$\\
Median \pratthree &$1.10\pm0.16$&$1.06\pm0.15$\\
Slope of  $p_{250}\mbox{ vs }p_{350}$ &$0.99\pm0.21$&$1.07\pm0.01$\\
Slope of $p_{500}\mbox{ vs }p_{350}$ &$1.03\pm0.05$&$0.87\pm0.01$\\
Slope of $p_{850}\mbox{ vs }p_{350}$ &$1.24\pm0.20$&$1.15\pm0.04$\\
$p(\lambda)$ linear slope $b_l (\times10^{-4})$ &\inlauralinearb&\outlauralinearb\\
$p(\lambda)$ power-law exponent $b_{pl}$&\inlaurapowerlawb&\outlaurapowerlawb\\
$p(\lambda)$ polynomial fit $b_{2p} (\times10^{-6})$ &\inlaurapolyb&\outlaurapolyb\\
$p(\lambda)$ polynomial fit $c_{2p} (\times10^{-4})$ &\inlaurapolyc&\outlaurapolyc\\
\bottomrule
\end{tabular}
\end{center}
\label{table:2pops}
\end{table}

\section{Discussion}\label{sec:discuss}
Figure \ref{allspec} summarized the polarization spectrum measured for Vela C, using all of the methods described in the previous sections. Also shown for comparison are the polarization spectra of other molecular clouds reported in previous work (see the caption). In combination, the measurements of the polarization spectra in other clouds are suggestive of a minimum at $\lambda\sim350$\,\um. However, the minimum is actually seen directly in only one individual source, OMC-1. The data for M17 are monotonic through 350\,\um~and all other sources lack sufficient wavelength coverage to show this V shape, or not, having polarization spectra that rise away from 350\,\um~on either the longward or shortward side but not both. In contrast, the BLASTPol data in Vela C are consistent with a flat spectrum (i.e. $p/p_{350}\sim1.0$) in the 250-850\,\um~range, regardless of the method of measuring the polarization spectrum.

Shortward of 350\,\um, the Vela C data cannot be directly compared to other clouds, because no previous measurements were made at 250\,\um. Measurements at 60\,\um~and 100\,\um~generally indicate a steep decrease in the polarization spectrum from 60\,\um~to 350\,\um, but the 250\,\um~measurement in Vela C does not follow this trend.

There are previous measurements of the spectrum at 850\,\um~by \cite{Vaillancourt2012} that can be compared with the Vela C measurement. There, the spectrum in other molecular clouds is again steeper than in Vela C, rising to a ratio of \pratthree~=\,1.6\,--\,1.7, compared to about 1.1 in Vela C.

The putative V-shaped far-IR decrease and submillimeter rise seen by other experiments have yet to be connected to a theoretical dust model. \citet{Hildebrand1999} and \citet{Vaillancourt2008} argue that the observed behavior is not consistent with a simple isothermal dust model, but requires multiple grain populations, where each population's polarization efficiency is correlated with either the dust temperature or spectral index. 

The \citet{Draine2009} models produce polarization spectra that rise from \pratone\,$\sim$\,0.9 to \pratthree\,$\sim$\,1.0\,--\,1.3, depending on the composition, shapes, and alignment of the grain mixture. However, their models apply to the diffuse interstellar medium (\av\,$<$\,4\,mag), not to dense molecular clouds such as Vela C. The \citet{Bethell2007} molecular cloud model is more relevant to this study and also predicts a spectrum that gradually rises from \pratone~$\sim0.9$ to \pratthree~$\sim1.1$.

\begin{figure}
\epsscale{1.25}
\plotone{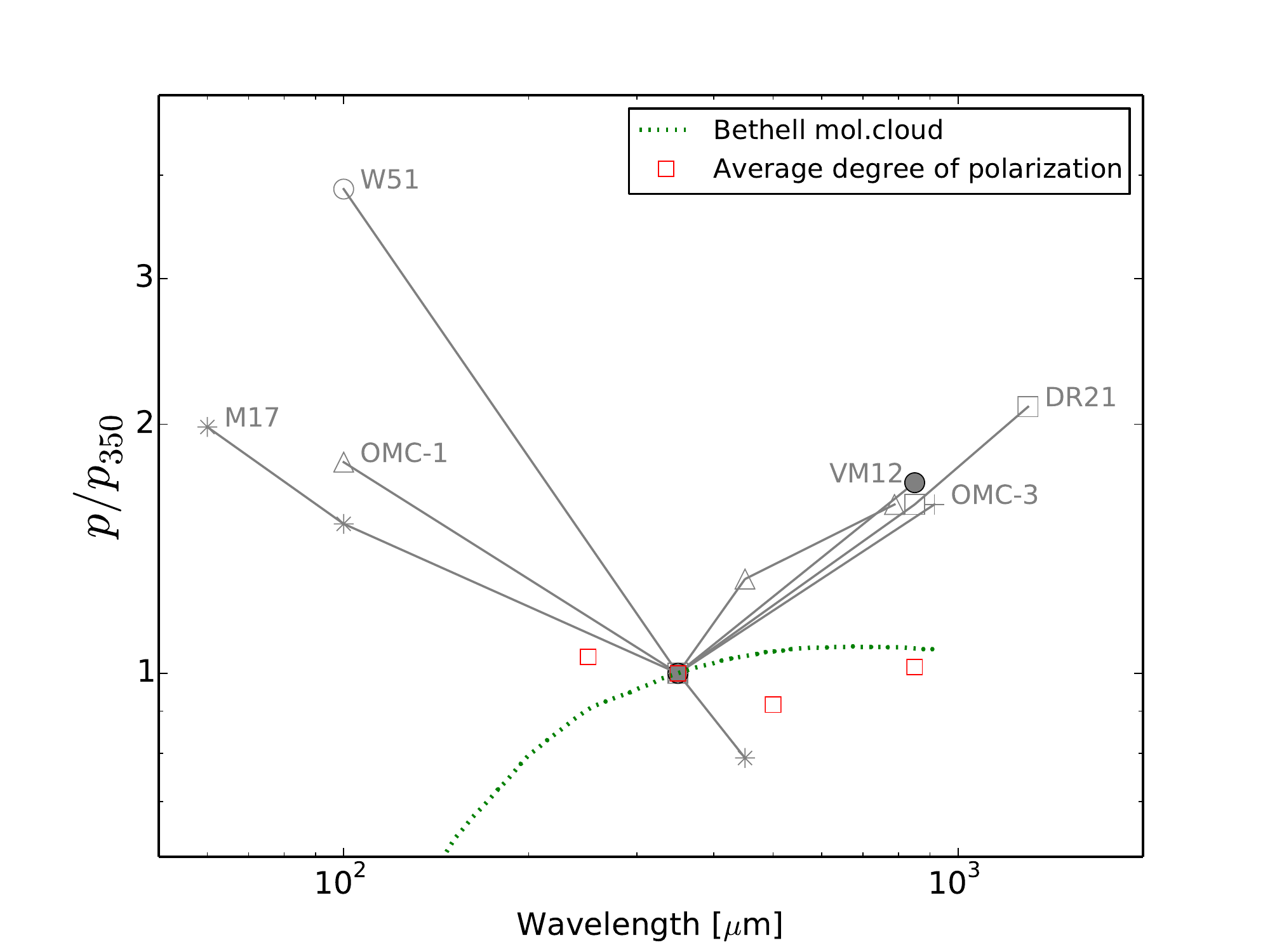}
\caption{Comparison with the predicted polarization spectrum from \citet{Bethell2007}, represented by the green dashed line. Red squares represent the total polarized fraction of the Vela C data, normalized to 350\,\um.\label{model}}
\end{figure}

A comparison of the Vela C data with the spectrum predicted by \citet{Bethell2007} over the 350\,--\,850\,\um~range is shown in Figure \ref{model}. The Vela C data are shown as the average degree of polarization at each wavelength (total polarized intensity, $P$, divided by total intensity, $I$), normalized to 350\,\um. The data were restricted to the region heated by the ISRF. Unlike the model, these Vela C data (as in Figure \ref{2pops}) show a slight minimum at 500\,\um, rather than a rise from 250\,\um~to 850\,\um. However, the data resemble the model more closely than the previous observations of other molecular clouds. While the model spectrum is fairly flat longward of $\lambda\sim300$\,\um, it falls precipitously at wavelengths shorter than what BLASTPol measured. Future measurements by experiments like HAWC+ \citep{Dowell2013}, a polarimeter for SOFIA operating at 50-220\,\um, would help to constrain the far-IR part of the polarization spectrum.

While \citet{Bethell2007} work under the assumption of starless clouds, in real molecular clouds there exist embedded stars that provide an additional source of photons. The part of the spectrum that increases toward the far-IR could be due to embedded sources both heating the dust grains and leading to a higher alignment efficiency through RATs \citep{Vaillancourt2012,Zeng2013}. 
The absence of a spectrum that increases below 350\,\um~in Vela C might be due to the early evolutionary state of the cloud. However, future work at high resolution could look for the effect of embedded sources in Vela C by measuring the polarization spectrum toward sightlines of known protostars.

A major difference underlying the data compared is that the Vela C data were obtained from balloon-borne and space-based observatories while the results from previous works (shown in Figures \ref{allspec}, \ref{2pops}, and \ref{model}) use a combination of data from airborne (12.5 km altitude) and ground-based observatories (longward of 100\,\um, the data are all from the ground).
Because they are observing through the Earth's atmosphere, these experiments are flux limited to very bright dense parts of molecular clouds. In contrast, the Vela C map probes a wider area of colder (11-15 K) dust in relatively quiescent regions. One way of quantifying the difference between the environment of the dust in Vela C and that in the clouds observed from the ground is by using the 850\,\um~intensity. For the data used by \citet{Vaillancourt2012} to study 17 molecular clouds, the median intensity at 850\,\um~is 637 MJy\,sr$^{-1}$, with an interquartile range of 300-1327 MJy\,sr$^{-1}$. In the Vela C data used here, the median intensity at 850\,\um~is 9.1 MJy\,sr$^{-1}$, with an interquartile range of 6.5-14.1 MJy\,sr$^{-1}$. 

The part of Vela C being radiated by RCW 36 is the most comparable to the bright regions of clouds observed by other experiments. However, in RCW 36-heated sightlines alone, the median 850\,\um~intensity is still only 20.2 MJy\,sr$^{-1}$, with an interquartile range of 12.4-25.4 MJy\,sr$^{-1}$. When the dust being irradiated by RCW 36 is analyzed separately, the various methods of measuring the shape of the polarization spectrum still give results that are consistent with a flat spectrum (Figure \ref{2pops}), in contrast to the V-shape.
It is worth noting that the region closest to RCW 36 was excluded from analysis (see Section 2), and the shape of the spectrum might be changed by adding those data points if such data were available.
 
Although the radiative environment of the dust in Vela C was quantified by $N$ and $T$, it is possible that more complex metrics are needed. For example, the RAT mechanism predicts that grain alignment is highly dependent on the anisotropy of the radiation field. Indeed, \citet{Andersson2010} and \citet{Vaillancourt2015} find a dependence of the polarization fraction on the relative angle between the radiation field anisotropy and the magnetic field direction. One might carry out such a test in Vela C using the peak in the RCW 36 intensity to define a relative angle of radiation anisotropy, but this has not been investigated given the low spatial resolution of the data.

\section{Summary}
\label{sec:summ}
A total of 459 measurements were made by \blastpol~in the Vela C molecular cloud, at 250, 350, and 500\,\um. These were analyzed with \Planck measurements at 850\,\um. The data were used to measure the polarization spectrum using several methods, including the median polarization ratios, the slopes of $p$ vs $p_{350}$ scatter plots, and fits to functions $p(\lambda)$. All methods indicate that the spectrum is quite flat, especially compared to the V-shaped spectrum suggested by previous observations in other molecular clouds. The polarization fraction remains relatively constant from 250 to 850\,\um~and does not depend significantly on the environment of the cloud, as quantified by $N$ or $T$ or by the source of irradiation (RCW 36 or ISRF). From 250 to 850\,\um~the spectrum in Vela C appears consistent with the predicted spectrum from \citet{Bethell2007} for a starless molecular cloud; measurements at shorter wavelengths would provide further constraints.

\acknowledgements
The \blastpol~collaboration acknowledges support from NASA (through grant numbers NAG5-12785, NAG5-13301, NNGO-6GI11G, NNX0-9AB98G, and the Illinois Space Grant Consortium), the Canadian Space Agency, the Leverhulme Trust through the Research Project Grant F/00 407/BN, Canada's Natural Sciences and Engineering Research Council, the Canada Foundation for Innovation, the Ontario Innovation Trust, and the US National Science Foundation Office of Polar Programs. Based on observations obtained with Planck (http://www.esa.int/Planck), an ESA science mission with instruments and contributions directly funded by ESA Member States, NASA, and Canada. C.B.N. also acknowledges support from the Canadian Institute for Advanced Research. F.P.S. is supported by the CAPES grant 2397/13-7. F.P. thanks the Spanish Ministry of Economy and Competitiveness (MINECO) under the Consolider-Ingenio project CSD2010-00064. P.A. is supported through Reach for the Stars, a GK-12 program supported by the National Science Foundation under grant DGE-0948017. Finally, we thank the Columbia Scientific Balloon Facility staff for their outstanding work.

\bibliography{submm_poln_spec}

\end{document}